# Cesno: The Initial Design of a New Programming Language

Ozelot Vanilla[1][0000-0001-8724-8292], Jingxiang Yu[2][1111-2222-3333-4444] and Hemn Barzan Abdalla[3][0000-0002-3637-0447]

[1] Computer Science Department, Wenzhou-Kean University, Wenzhou, China
shuih@kean.edu
[2] Computer Science Department, Wenzhou-Kean University, Wenzhou, China
yujingx@kean.edu
[3] Computer Science Department, Wenzhou-Kean University, Wenzhou, China
habdalla@kean.edu

**Abstract.** Programming languages are incredibly versatile, enabling developers to create applications and programs that suit their individual requirements. This article introduces a new language called Cesno, designed from the ground up to offer an advanced, user-friendly, and easy-to-use programming environment. Cesno's syntax is similar to other popular languages, making it simple to learn and work with. It incorporates features from other languages, such as syntactic sugar, a built-in library, support for functional programming, object-oriented programming, dynamic typing, a type system, and a variety of function parameters and restrictions. This article will explore the design of Cesno's grammar, provide a brief overview of how Cesno processes and compiles code, and provide examples of what Cesno's code looks like and how it can aid in development.

## 1 Introduction

Programming is the process of utilizing logic to execute specific tasks and functions on computers. These logical operations are implemented by means of specific programming languages, which comprise a set of instructions and commands written in a distinct manner to create a unique program and direct the computer to execute it. Programming languages also regulate the way electronic devices communicate with each other, such as robots and peripheral devices like printers and other smart de-vices, as well as permitting humans to communicate with machines. Even though various programming languages share many similarities, each language has its own distinct structure and a specific set of keywords to develop a certain program.

Also, a programming language is a set of commands written according to a set of rules determined by the language itself. These commands go through several stages before being executed on a computer. Programming languages are divided based on their proximity to human language into high-level languages (closer to the language humans understand) such as C and Java, and low-level languages (closer to machine language, such as Assembly language). Some languages are designed to work on specific devices, such as a computer or a central processor (CPU). In these cases, the company that produces the device provides a user manual that includes the commands that can be executed on it. Other more general languages, such as Java, work independently of the type of machine, working within a virtual machine.

Drafting, Syntax in programming languages are rules that de-fine the correct way of writing symbols and vocabulary within a program based on the language used. There are specific symbols and reserved words for each language, such as the word “IF”, which is used in a specific way. Some languages consider the use of lowercase letters similar to the use of uppercase letters, while other languages are the opposite. Semantics refers to the correct usage of symbols and vocabulary to form a code sentence in accordance with the Syntax rules. Usually, these sentences are executed sequentially, and the sentence being per-formed at the moment is valid. Types are data that have specific properties that are checked by the compiler, and an error is shown if a form of data is assigned to an incorrect type. Static type checking and dynamic type checking are two methods of examining types. Libraries are collections of utilities and properties provided in a language for program development. Previously, libraries weren't seen as an essential part of languages, but with their tremendous development, using these helpful tools for writing more efficient programs has become a necessity.

## 2 Related Works

### 2.1 Development of Programming Languages

The development of programming languages has been ongoing since the first computer programming language was developed in the 1950s. Since then, numerous programming languages have been created, improved and adopted for various purposes. Early programming languages such as FORTRAN and COBOL were developed to facilitate scientific computing and business applications; compared to assembly language, they get rid of the limitation of different architecture (become architecture independent) and have a more human-readable syntax design [1], [2]. Later, more complex languages such as Pascal, C++ and Java were created to provide more powerful programmability and provide support for object-oriented programming. In recent years, numerous new programming languages have emerged, such as Python, Ruby, Rust, and Go. These languages have been designed to enable developers to create robust applications while also making development faster and easier [3], [4]. The development of programming languages continues to evolve with the emergence of new technologies and trends. For example, popular languages such as JavaScript are being used to create web applications, while languages such as Swift are being used to develop mobile applications. Furthermore, the development of programming languages has been greatly aided by the availability of powerful tools and libraries. These tools and libraries make it easier to develop applications in a variety of languages, while also providing support for popular frameworks such as React and An-gular. The trend of programming languages is to make it easy to learn [5], develop and maintain, and let the people not in IT have a chance to program [1].
Although there are many programming languages, not all of them can be well known over decades [6]. New programming languages emerge to meet the new requirements and platform [7], [8], but new languages may lack features which are available in the previous programming languages, and people need time to get familiar with new languages [8].

Most programming languages can be divided into static and dynamic type systems. For example, Rust or C++ use static type, while Python or TypeScript use dynamic one. Software developers may choose static typing because of its reliability of higher maintainability [9]. However, there are also arguments that declaring static type may be time consuming and hard to change return type [9]. In contrast, with easy but clear syntax, dynamic typing can also help developers to write less and have higher productivity [10]. To combine the advantages of both, some programming languages like Groovy provide both of them [10].

Programming language design should also focus on the educational aspect, such as the cost of learning, and friendliness to non-programmer. A paper which introduced a graphical language [11] suggested that programming should be interesting, and be able to let people get some skills of programming even if they may not be programmers.

### 2.2 Some Drawbacks of Existing Programming Languages

People may be confused by the formality of writing code in some programming language [12]. Java is an object-oriented programming language, and it forces to write content like "public class" and "public static void main" for every program, even though the program might only print "Hello World" in the console, with no argument received and used. C and C++ also need to include header files even for standard input and output [13], or perform the power function.

When it comes to learning, current programming languages may not be easy for first-time learners – some people might misunderstand some concepts if they learn by "looking at sample code" and then "try to find rules and combine the function" mode (top-down approach) [14]. But for the bottom-up approach (first learn grammar, then apply), a C program includes too many concepts such as usage of "&" and "%d" for getting standard input. Also, C or C++ may discourage first time by implicit type convention, different behaviour on different compilers, or out-of-bound array indexes [13], and it is designed to be fit for experienced programmers, but not guaranteed to be simple and reliable [15].

Programming languages may offer some features with confusing form. For example, the operator "sizeof" in C++. Although it looks like a function call, the compiler will substitute it while compiling. For example, "sizeof(i++)" equals to 4 if "i" is an integer. However, calculation "i++" isn't executed. Because the compiler just simply checks the variable's type and puts a number on there. The directive "#include" in C++ can be also a burden for programmers. To make a class usable in one file, user must use "#include" to import them. While other programming languages had already moved to the module system, C++ maintained this 30 years old technique for importing [16]. Also, "#include" is malfunctioning when there are two files including each other, or one header file being includ-ed twice or more. To overcome these problems, user have to write code such as “#ifdef ANIMAL_CLASS_HEADER” [16] or “#pragma once”, which is not related to the logic and target of programming tasks (that means, these code segments are not ought to exist).

In addition, programming languages may offer too many restrictions on expressing, which may cause negative effects, it might be inconsistencies in expressing the same thing under different context, while according to daily logic, it should work. For example, Java only allows the array initialised by literal like “{1, 2, 3}”, and does not allow any other usage of that literal [15]. Also, Java forces users to write import statements on the top of the file, but this restriction may reduce the freedom of importing names only inside one function, which affects the expressiveness [15].

Also, Programming languages may lack features to let users write understandable or maintainable code, like JavaScript is based on the dynamic and weakly-typed type system, and it uses prototype as a way of inheritance; these features may confuse programmers and they must learn a lot of features of JavaScript to make the code maintainable [17], or to say, the language itself is too flexible to write or use programs easily in a big project [7], [18]. In JavaScript, the implicit convention of type is common when the operand is not the same type, which may be hard to understand. For example, both “`"3" == 3`” and “`3 == "3"`” returns true; but both “`"true" == true`” and “`"false" == false`” returns false. “`"1n"`” could be parsed to integer using “`parseInt`”, but “`"1n"`” is neither equal to “`1`” or “`1n`”; however, “`"1" == 1n`” is true.

### 2.3 Goal of Cesno

To solve these problems described in the previous section, the proposed language should have these characteristics: not so heavy to learn to write simple programs, keep consistency and coherency on the design of syntax, allow  to code more freely by less restriction and formality, offer proper grammar sugar to make the code efficient and understandable, be able to be non-domain-specified and offer enough features for  who need to write high performance and optimised program.

The proposed language, Cesno, is aimed to combine the advantages of different programming languages together, while the grammar is not messed-up. Cesno is a C-style programming language, and it borrows a lot of ideas from languages like Python (e.g. handful builtin functions, positional-only and keyword-only-argument), TypeScript (e.g. type manipulation, arrow function, function declaration), and Rust (e.g. evaluate value of structures like match statement, enum member that carries more information). The expected outcome of Cesno will be (for the detailed description, please see [1]):

1. Precise grammar with unified specification.
2. Supporting different paradigms like Object-Oriented and Functional style.
3. Easy to learn, with convenient feature, without being limited by stereotype.
4. Ability to adapt for simple or complex usage.
5. Zero cost abstractions.
6. Pay for more memory and speed cost if using extra features.
7. Open to customization.
8. Not being limited by specific domain, architecture, or platform.
9. Up-to-date features, learn from others, learn from users, learn from suggestions.

# 3 Implementation Model

This part, demonstrate the basic grammar model of Cesno, in addition, an implementation of MCU for the brief basic grammar model will be given.

## 3.1 Basic Grammar

This part describes the basic design of syntax and components in Cesno. The sections below are categories of statements in Cesno with syntax design and examples for each part.

**Components of Cesno Code.** This part demonstrates the construction of Cesno code. The code in Cesno is built with the *language gadget*. Everything in the code can be seen as a *language gadget*. The smallest language gadget includes the literal, identifier, or punctuation, while the more significant language gadget includes a function or class definition.

***Statement*** is the grammatically correct combination of several language gadgets, which is one important structure of language gadget. There are mainly three types of statement:

1. Identifier declaration, assignment, and management.
2. Function or method call.
3. Structure (like *if-else* in most languages)

A statement is also a language gadget, so it can include one or more statements, also. Additionally, a statement may not be functional (that means it can do nothing) – as long as its grammar is correct, the statement is established. Many language gadgets have expected tokens defined before, and it waits for these tokens to finish the statement.

The ***keywords*** in Cesno are mostly defined as a language gadget, it expects tokens, and become available as a keyword when the expected token is fulfilled. Therefore, function can use the keyword's name as its name, since the language gadget "function" is expecting a name before the parameter's tuple, and the keyword is not effective in the expectation of "identifier".

**Identifier declaration and assignment:** Identifier declaration has type name and identifier only, while assignment has an assignment operator and right value. The following rules of identifier declaration and assignment are as follows. This will make the identifier declared.

1. `MODIFIER TYPE_NAME IDENTIFIER ;`
2. `IDENTIFIER: MODIFIER TYPE_NAME ;`
3. `MODIFIER TYPE_NAME IDENTIFIER = DEFINITION_WORD EXPECTED_TOKEN ;`
4. `MODIFIER TYPE_NAME IDENTIFIER ASSIGNMENT_OPERATOR RIGHT_VALUE ;`
5. `MODIFIER TYPE_NAME IDENTIFIER DEFINITION_BODY ;`
6. Shorthand notation, starting with "`let`", "`const`", or "`auto`".

For the third rule above, the "`DEFINITION_WORD`" (which will be talked about later) should be corresponding to the declaring identifier's type. For the fourth rule, the "`ASSIGNMENT_OPERATOR`" can be "`=`" (assign), "`+=`" (plus-assign) or something familiar.

Example 1: declare variable “a” as a “int”, declare and assign “b” as a “int” with value 10. After that, assign “b” with its value plus 20, and give that value also to “a”.

```
1       int a ;
2       int b = 10 ;
3       a = b += 20 ;
```

Example 2: define a function with overload, first one can add an int type variable with 1, second one can add a float type variable with 5. Let the return type be inferred by the return statement.

```
1      function addNumberBasedOnType(int integer_num)
2      {
3          return integer_num + 1 ;
4      }
5
6      function addNumberBasedOnType(float float_num)
7      {
8          return float_num + 5 ;
9      }
```

**Function or method call:** It has either function or method (with instance it belongs to) name and a tuple of arguments. This type of statement is usually nested in other statements.

1. `FUNCTION_NAME(ARGUMENTS) ;`
2. `function(PARAMETERS){ FUNCTION_BODY } (ARGUMENTS) ;`
3. `VARIABLE_OR_TYPE.METHOD_OR_FUNCTION_NAME(ARGUMENTS) ;`

Example: define a function which receives two numbers, and print the sum value. Define a partial function which will print the received value plus one, call it with “1” (partially applied function).

```
1      function addThenShow(int a, int b)
2      {
3          print(a + b) ;
4      }
5
6      let printValueAddOne = addThenShow with b = 1 ;
7      printValueAddOne(1) ;
```

**Structure:** It starts with a word which is registered into the compiler (including macro), then it will follow the defined rule to receive the following tokens. It contains two parts – definition and flow control.

Definition comes with a *definition word*, an identifier’s name and *definition body*. The definition word is one word that indicates what will be defined (for example, whether you are going to define a class, or an enum). The identifier should be written for the receiving of the definition body. The definition body (which is EXPECTED_TOKEN below), which is the concrete definition offered to the identifier, varies according to the definition word (for example, the definition body of class and function are different). At the end of structure, users can append names for instances, if the previously defined structure can be a type, and users want to create instances with that type immediately.

1. `DEFINITION_WORD IDENTIFIER EXPECTED_TOKEN ;`
2. `DEFINITION_WORD IDENTIFIER EXPECTED_TOKEN INSTANCE ;`
3. `DEFINITION_WORD EXPECTED_TOKEN ;`

Example: define a class "Book" with string member "title" and string array member "author", all members are read only. One two parameters constructor for the class.

```
1    class Book
2    {
3        readonly string   title  ;
4        readonly string[] author ;
5
6        constructor(string title, string[] author)
7    }
```

Flow control comes with a *flow control keyword* (including keyword provided by macro), which changes the running order of code, letting them may not be run from top to bottom. The `EXPECTED_TOKEN` is the code controlled by the flow control keyword.

1. `FLOW_CONTROL_KEYWORD EXPECTED_TOKEN ;`

Example: if the code received a value bigger than 10, print a sentence "bigger than 10"; if not, print "smaller or equal to 10".

```
1    if (int(input()) > 10)
2    {
3        print("bigger than 10") ;
4    }
5    else
6    {
7        print("smaller or equal to 10") ;
8    }
```

**Identifier management:** It is the operation that affects the behaviour of the identifier or the functionality behind, including memory management and concurrent processes. For example, import statements define names that refer to the imported package or function, and that defined name (identifier) is not able to be used in the same scope (otherwise there would be name conflict). Another example can be the deletion of the variable, after delete one variable, the memory is freed, as well as the identifier (it can be used again after the deletion).

Example: delete an `int[]` variable after definition.

```
1    int[] a = [10, 20] ;
2    delete a ;
```

**Other types of statements:** It contains statements such as those that can be seen as grammar sugar, user-defined macro, or statements coming from another language. For example, Cesno allow user to create a code block that applied to some modifier as showing below:

```
1       export
2       {
3       function somethingToExport1() { } ;
4       const something_to_export_2 = 0 ;
5       }
```

Above code, the function and the variable share the same modifier "`export`", so users can move them in one code block after "`export`". This feature can also be used in class (e.g. access modifier "`public`").

**Literals**: Literals are the smallest language gadget in the code, and it is the direct value for representing instance (since Cesno is Object-Oriented, so the `int` or `float` are also class). Literals are able to be called methods or functions from, but themselves are constant. For detailed pre-defined literal definitions and descriptions, see [1].

**Control Statements.** For the statement with loop feature, like "for", Cesno uses the *iteration indicator* to indicate the variable used in the loop. Users can use indicators like C style programming: initialise, run-condition, after-loop action (`(int i = 0; i < 10; i++)`). Also, users can use "in-iterate" to get a pre-defined iterator variable (`(let c in "Hello")` will go through the string character by character, not splitting the ascent). It is also possible to define a new iterator indicator, for example, users can define an "of-iterate" like JavaScript.

Unlike traditional C style programming language, Cesno allows users to define multiple different types of iterator variables at the same time. Each variable used in the loop will go "one-step-after" after one iteration. When one of the variables runs out (end in its iteration), it will become `undefined` in the next iteration. When all variables run out, the iteration will come to an end.

Cesno also allows users to write "`break indicator`" or "`continue indicator`" to break, or continue outer-loop inside inner-loop, avoiding using tags. The indicator will be one of the variables inside of the outer-loop's iteration indicator.

It is possible to add "`then`" and "`else`" clauses after a statement with a loop feature. The code in the "`then`" clause will run if the loop is not ended by a "`break`". If so, the code in the "`else`" clauses will run. The control statement has evaluated value as well, the evaluated value of the last run statement will be that value.

**Basic Usage of Operators.** This part introduces the basics of operators in Cesno. It is possible to define new operators in Cesno, as long as the new operators are not parenthesis, dot, comma, semicolon.

**Elementary arithmetic operators and remainder**: Addition ("`+`"), subtraction ("`-`"), multiplication ("`*`") are working almost the same as mathematics (ignoring overflow, underflow, or precision), while division ("`/`") depends on operands' type. The sign of the remainder's result follows the left operand. If users want to use the modulus operator, it is able to be imported. Increment ("`++`") and decrement ("`--`") operators are almost the same in other languages which support them (and by default, atomic in Cesno).

**Logical operators and bitwise operators:** Unlike most C-style languages, Cesno uses "`and`", "`or`", "`not`", and "`xor`" for logical operators; also "`bitand`", "`bitor`", "`bitnot`", and "`bitxor`" for bitwise operators. For bit shifting, Cesno uses "`bitshl`" and "`bitshr`" for signed bit shifting, "`bitushl`" and "`bitushr`" for unsigned bit shifting.

**Validation operators:** This type of operators includes such as equality-checking operators, nullish-checking operators, and type-relation checking operators.

Equality-checking operators works like most C-style languages, such as "`<`", "`>`", "`<=`", "`>=`". Operators "`==`" and "`!=`" (by default, it returns "`not`" result of "`==`") can be defined to test whether two variables are equal or not, while "`===`" compares if two variables are using the same address. Operator "`~=`" is only predefined for float numbers, to test the equality approximately (since it is harmful to test two float numbers with "`==`", if users try to use "`==`" to compare the float number, there would be a warning).

Nullish-checking operators test if a variable is null, undefined, or even falsy. Operator "`??`" tests if the nullable variable before is not null or undefined, while operator "`???`" also checks if the variable is not falsy. Users can also define the rule of checking whether a variable is falsy. For example, user can implement trait "`HasFalsy`", and define constants that is considered falsy. The detailed example and code could be seen at [1].

Since some operators are able to be overloaded, operators that are the same in text may have different names and usage. For example, "`+`" sign can be "addition" or "append", depending on the type of the operand. If the operands can be inverted and not affect the final result (commutative), it is "addition". If the order matters, it is "append". This design will be useful to tell the programmers which trait they are implementing, since the code and functionality will be different.

For the evaluation of operands and operators (in this case, ignoring operators that have three or more operands, currently, original Cesno only contains one ternary operator "`? :`" at now), Cesno will do these in order:

1. Find if the type of operand (unary) / left operand (binary) has a definition of that operator. For example, left-add operator for int: `operator (+)(int right)`.
2. If binary operators, find if type of second operand has definition of that operator. For example, right-add operator for int: `operator right (+)(int left)`. Cesno always first checks the left hand side, then the right hand side for the binary operator.
3. Find if there is an imported operator that fits the operand(s). These definitions are not defined inside the type, but defined globally. For example, there is no `operator (+)` for `string` and `int`, but if define that and let this definition available in some scope, it will be able to add an `int` to a `string` in those scope.
4. Find if the original Cesno has the definition of that operator. But in most conditions, Cesno already defines operators for the basic types.

**Basic Containers.** This part introduces the basic containers and its representation. Basic containers are not suggested to be re-defined by the user.

**Tuple:** Tuple saves one or more non-empty (but nullable) elements as constants. It has a type that combines each type of the element. For example, a tuple "`(int, 0, "s")`" has a "`(type, int, string)`" type, while a "`(0, 0)`" has a "`(int, int)`" type. Unlike a struct, the value in the tuple is not-named, and it is queried using index.

If it does not have any element, tuple will be seen as "`void`" typed, and it has only a zero-size "`empty`" element, which uses no space in the memory.

**Array:** The array is a linear container that stores elements continuously, guaranteeing the constant time complexity for searching. It has a type of "`ValueType[]`" (this is the shorthand notation for “`array<ValueType>`”), and its element must be “`ValueType`”, or can be seen as it, or is one of the “`ValueType`” if it is a union type. Arrays in Cesno are flexible to length, because users can push/pop elements from head to tail. However, it might be time consuming to perform these operations, since arrays need to save the elements sequentially inside the memory.

**List:** The list works like an array, it looks like a linear container, but it may not have a constant time for searching the element. List can also push/pop elements from head to tail, but it can be done more efficiently since lists may not store elements in the memory sequentially.

**Sequence (or Seq):** The Sequence is the fixed-length array in Cesno. It works like an array because they are both linear containers. The sequence type not only contains types of internal elements, but also the length of itself (`sequence<type EleType, usize SelfLength>`, or `EleType[SelfLength]` like `int[3]` for shorthand notation). Hence, the sequence that has different length is not the same type.

**Dict (or Dictionary):** The Dict saves pairs of keys and values. Keys must be different in the dict, while the values are not. The keys of dict are not guaranteed to be ordered, but there is an ordered version of it. To read the value of dict, use key as the index.

**Set:** The container which contains non-repetitive elements. Set is not guaranteed to be ordered, but there is an ordered version of set. Set is not indexable, to read every element of set, use iteration. Set has a type that can be seen as a void-typed value dict, with different functions and methods (“`type set<type EleType> = dict<EleType, void>`”).

To avoid confusion, if a tuple, or a set, only contains one element, users may put a trailing comma like “`{0,}`”, or use the constructor instead.

**Function and Method.** Function is a process which accepts some arguments, does some action, and in the most time, gives back some results. Method is similar to function, but it is bound to an instance, able to read or write the instance’s data. Function and method are objects in Cesno, they can be assigned to variables or passed as arguments.

To define a function, it needs to write “`function`” first, and write some identifiers’ name for its return type and name, parameter’s tuple, and definition body. It is not forced to write the return type explicitly, since there is type inference.

For function declarations, there can be an *overload*. Functions with different signatures will be seen as different definitions. To make different signatures, different parameter types or limitations for a function can be applied. Like Python [19], it is able to use “/” or “*” to make parameters passed *positional-only* or *keyword-only*. Users cannot use “key with value” to pass value to a positional-only parameter, nor pass value to keyword-only parameter without its key.

It is also possible to use “`TypeName...`” as a *variable length type* to indicate that the function can receive 0 or more parameters passed as a tuple. Unlike Java, there can be multiple variable length types in one parameter list, as long as the compiler is able to distinguish which argument is passed to which parameter. For example, “`(string... texts, int... integers, string str)`” is acceptable, since there is an obvious difference of type between the first, second, and the third parameter. “`(string... texts, string word, string... strs)`” is not suggested because until user use

"key and value" to tell the compiler which value is passed to the second parameter `word`, the compiler cannot decide how to separate arguments that user has inputted into these parameter; If the parameter `word` is positional-only (`(string... texts, string word, /, string... strs)`), this parameter list will not be acceptable, since there is no way to tell the compiler how to separate arguments.

Users can also use a question mark like "PARAM?" to show that the parameter may not be passed with value from the user. However, if two functions' signatures are effectively equal, there would be an error. For example, function signature "`(int a, int b = 10)`" and "`(int a, int b?)`" are effectively equal ("`int b = 10`" is shorthand for "`int b? = 10`", or "`int b?`" with in-function checking-and-assigning).

There is a feature called *parameter constraint* in Cesno, which can help narrowing the argument's type, check if arguments fulfill some prerequisites, or make the code more elegant. By adding a colon after the parameter name, and writing *constrainers* concatenated with "`&`" or "`|`" (which means "and" and "or" for condition), can narrow the acceptable argument, or check if the precondition is fulfilled. Constrainer is the condition to check with the argument. It can be a modifier check (like if the argument is a constant), a type check (like if the argument is a linear container), or a bool check (like if the argument is bigger than 10). Constrainer helps the compiler check if the argument is problematic as soon as possible. Modifier check and type check can be done statically at the most time, while bool check may also be done before the runtime (given a const expression, `constexpr`). A mock example for each constraint looks like this:

```
1    function (Container a: const & Lengthwise & (a.length > 2))
2    { }
```

In the above example, the argument passed should be a constant, lengthwise, and have a length bigger than 2. If any of them is violated, the compiler would not let it compile. Although the bool constraint may not be working at compile time if a variable is passed, it will be checked before the function runs, and give a `ConstraintNotFulfilledError` if the argument does not satisfy the constraint.

It is the same to write "`(int a)`" and "`(a: int)`" in Cesno, since the second representation means "an `any` typed `a` has an `int` type constraint", which is equivalent to the former which means "`int` typed `a`". However, modifiers have different meanings before and after the semicolon. If you want to receive a value as a constant in the function body, you write "`(const a)`" (like defining variable); if you want the argument to be a constant when passed into the function, you write "`(a: const)`" to indicate that this function only allows constant to be passed (since there is a modifier constraint).

There might be ambiguity for understanding the identifiers between the "`function`" and the parameter's tuple. If there is no identifier, it is an anonymous function with type inference; if there is one identifier, Cesno will use this identifier as the function's name; if there are two, they should be the return type and function's name. When users want to create anonymous functions with checking on the return type, they might move the return type afterward, as a constraint for the function:

```
1    function(a): returns int, a: int
2    { return a + 1 }
```

Since function is an object, it can also be passed as an argument. If a function is passed as argument, there will be an assignability check of functions. For example, a `map` method of array has a type constraint as follow (“`->`” is type annotation for function, the tuple before is parameter, and the type after is return type):

```
1     method <type MappedType> map(
2         mapper:
3             (element: EleType,
4              index?: int,
5              this_arr?: EleType[])
6              -> MappedType
7     )
```

This method requires the argument for `mapper` to be a function, which at least has one parameter `element: EleType`. If a function like `int(value, *, int ratio = 10)` constructor is passed, the parameter that is keyword-only will not be seen as the parameter of required `mapper`. Thus, problems like passing `parseInt` to `map` in JavaScript and getting unexpected results will not happen in Cesno – as long as the parameter is set to be keyword-only. If users want to change the default value of a keyword-only parameter, they can use “`with`” to define a partial function. For example, if users want to map a string array to int, but using ratio equals to 16, users can write `["0x1", "0xE"].map(int with ratio=16)`. The statement with “`with`” is a shorthand for defining:

```
1     // "int with ratio=16" can be seen as this function:
2     (
3     function intConstructorWithRatio(int ratio_value)
4     {
5         return
6             function (value)
7             {
8                 return int(value, ratio=ratio_value)
9             }
10    }
11    )(16)
12
13    // Which is finally equivalent to this:
14    v -> int(v, ratio=16)
```

To make the user-defined function work with the original Cesno function/method well, parameters that are “not the most important part” (and works more like a config to function) should better be set to keyword-only. For example, read text from a file. The file path is the most important part, while the encoding is a config that tells the function how the file would be read and understood.

**Type Manipulation.** This section introduces the idea of *type* and the operation available on types (*type manipulation*). The idea of type manipulation borrows a lot from TypeScript or Haskell.

Cesno’s type acts like most other programming languages, it offers the information on how a typed-variable can be used. Type can be seen as a set, or a rule of how data will be constructed. A variable with type A can only be either the member that A has, or the data that follows A's rule. For example, a variable with type `int` should follow the rules of `int`, it can only be an integer inside the `int`'s range; it is also correct to say that a variable with type `int` should be one of the members in the set of `int`, but this way of understanding may be hard for the compiler, to define or understand types allow almost infinity members (e.g. such as real number, string, or complex data structured by class).

For the naming convention of types, Cesno uses all lowercase letters for the type that should not be considered a combination of other types. For example, a `string` type (depending on different implementations on different platforms) will probably be the combination of types that is more "primitive", but Cesno does not want users to consider that when coding. If a user wanted to design a "rune" class that is the basic character of an artificial language in a game, that user would probably name it like `rune`. For the type that could be highly considered as a combination of data members, it names using the upper camel case like `FireRod` (which is probably a combination of value of attack, magic, runes possessed, and cool down time). For generics parameters, use descriptive names in upper camel case, no matter the actual type of generics, like `KeyType` (type generics) or `ElementSize` (integer generics, probably typed as `usize`).

Because type is not necessary to only be the rules, but can also be the set of possible values, a constant like "`false`" or "`"off"`" is also legal to be a type, called *literal type* – it describes a set that only contains one element.

There are some operators that produce new types from existing types. Operator "`|`" is used to combine the possible types together to make a new *union type*; for example, "`"off"|int|undefined`" is a type that allows a string with content "off", `int` type, or let the variable to be `undefined`. Operator "`&`" is used to get the intersection of the types; if two operands have some same members (variable, function, method), they will be extracted. Operator "`+`" is used to create a new type that contains two operand types' members, while operator "`-`" deletes the member of the second operand, from the first operand. The operand may not strictly follow the name in mathematics, it works more like the operand of list.

When a union type variable is passed to function, each type in the union type will be checked if it is able to be the acceptable type, and the return type will also be the union type of the return types. For example, the operator "`??`" returns `true` for type `object`, and `false` for `null` type or `undefined` type. If operator "`??`" checks variable with type `int|string`, it will always return true, since "`(returnof operator (??)(object)) | (returnof operator (??)(object))`" will be "`true | true`", and result is "`true`"; if it checks variable with type `int|null`, it will return "`(returnof operator (??)(object)) | (returnof operator (??)(null))`", which is equivalent to `bool`, "`true | false`".

Cesno is null-safe (and also undefined-safe). Unless user declared a variable can be null or undefined (e.g. `string|null`), it is not allowed to assign null or undefined to that variable. Since nullable or undefinable variables are union type, its value cannot be directly used (must use "`match`", null-undefined coalescing operator "`??:`", or other ways to narrow the type).

For type conversion, Cesno will not do it implicitly. Statements like "`if (10) { }`" are illegal in Cesno, since "`10`" is by default an "`int`" type, while "`if`" is expecting a "`bool`" type after it. For assignment or passing arguments, until the overload of assignment operation is defined, there will also be no implicit conversion.

Unlike TypeScript, Cesno also allows users to use some function to manipulate types, but not only utility types. Functions like "`(constexpr string s) -> s + "!"`" (append an exclamation mark after the given string), which is possible to get a value at compile time if the parameter is also constant, can be used in type manipulation. Therefore, users may reuse some functions, and avoid using utility types as a function (for example, a type that capitalises string may be found for a type that summons getters and setters method like "`getUserList`", actually it works like a function more. If using a function, some method like "`toUpperCase`" can be called, avoiding repeating these methods and making types like Intrinsic String Manipulation Types in TypeScript [20]).

**Other Part of Design.** For the design of class, generics, module system, code evaluation and compilation, built-in library, please refer to [1] for more information.

## 4 Drawbacks & Future Work

This article describes the design of Cesno, and the possibility of having such a model of language. Yet, we have not completed all the design, and the basic structure of syntax may still change. Also, the infrastructure like compiler, language library, package manager are not implemented. Hence, there is a lack of data of speed/memory usage of Cesno.

Some features are not discussed. Concurrent programming in Cesno is going to be implemented at a language primitive level, in order to make it able to fast-implement basic concurrent programs using structures like "`if`" statements. Also, how Cesno detects the end of a statement is not completely described.

In our future work, we will focus on several essential points.

**First**: There would be more content on how the language builds itself – how each word and symbol is recognized, how high-level languages' features (like class) map into low-level languages and become machine-optimised. It is important to answer these questions to make the programming language customizable, extendable, and allow users to do meta programming based on that system. Also, the built-in function needs to be designed to be more powerful. For example, how to implement "`format="${array; i32; /\d+(?:,\s*\d+)+/; catch=/\d+/}"`" in the "`input`" function mentioned above, letting the input to be checked, and giving back an array type instead of string. This requires the programming language to have a responsible meta-programming system.

**Second**: One selling point of Cesno will be the ability to customize the language easily, since the way the compiler compiles will also be represented as human-readable language, not hard-coded into the compiler. One approach is to use some keyword-like word to trigger the matching of pattern. For example, the "for-loop" has a trigger of "`for`", waiting for at least one *iteration indicator* (like "`(int i = 0; i < 10; i++)`" or "`(const x in [1, 2]`") and a code block. The trigger "`for`" can only be available if it is allowed in the "scope" – whether the trigger word is supposed to be here. If a "for" appears after "`method`" and before a pair of parentheses (`method for()`), it will be considered a name of the method, since "for-loop" structure is not supposed to be at the position of name, so the "for" is not considered as a trigger of "for-loop", but a identifier. However, since the code analysis mode is "expecting tokens", it is hard to write the compiler when the overall design is not finished. Design for how to implement functions that let the compiler read user-customised macro, literals, or statements also requires a strong will also have an impact on the compiler's time complexity.

**Third**: Cesno also wants to implement a strong type system, which defines how types form themselves, and how they are translated into assembly language. The first part is how value is considered to be some types – is it a value that inside a countable set (like `int` range from -2^31 to 2^31-1, or literal types like `"off"` (the variable with this type can only have one possibility of value, which is `"off"`)), or is it zero or more same-type elements' combination (like `string` (raw string is combination of `char`) or `array`), or is it combined with a finite number of members with some types (like builtin or user-defined class). These categories are useful in representing type's definition and value acceptable in a human-readable way inside the programming language. The second part is to disclose the function, method and its implementation detail of very basic and simple types such as `bool`. For example, some languages use "primitive type" to describe them, and omit the language-level implementation (like Java's int type or double type); also, it is not possible to call function or method directly from those types, until they are boxed. However, Cesno will not use these boxed types, and offer a post-implement feature for the type defined as mentioning value acceptable.

Another part to design is type compatibility. This is related to how Cesno checks the assignability and how the assigned variable works.

There is the difference of nominal type and structural type in programming languages, languages like Java or C# use nominal type, while TypeScript uses structural subtyping [21]. In this article, only nominal type (defined by class) is discussed. However, there could be structural types defined by "`struct`" in the future. If the necessity of structural type is proved, this system will be implemented. For nominal types, the assignability depends on whether they are the same, or have an inheritance relationship, since for types possessing the same members, they might have different meanings. For structural types, as long as required members are given, the assignability establishes.

Function assignability is different from other variables' assignment. As the previous section "Function and Method" describes, Cesno allows optional, positional and keyword arguments; when check the assignability, there will first be the check of arguments' number and name, whether they match the required function. After that it is the type check, the argument should be contravariant and the return type should be covariant to keep the assignability. Besides, Cesno has not discussed the strategy of picking appropriate function/method when there is overload of functions.

For generics, its design is not fully described in this article. Unlike some programming languages which only support classes (or types) as the generics type parameter [22], Cesno's generics will have less restrictions. For example, the basic container "sequence" will have a length parameter which is typed usize (`class sequence<type EleType, usize SelfLength>`). The core idea of generics is: the value that makes one type a different type from others should be put into generics' parameters list (e.g. element type of array). In order to accomplish this goal, the generics should be able to accept type, enum, function, or value with concrete type. The implementation of generics is not decided in this language design proposal (use template like C++, or use monomorphization like Rust, or allow both of them), and for generic parameters that is known until runtime, there is not a proper solution for assignment, type checking and so on.

**Fourth**: ability to adapt language features that are considered as basic features of a programming language. The ownership system is a basic feature that cannot be violated in Rust (or the program will not compile) [23], which offers higher performance and safety, for example, avoiding the unexpected change of shared states [4]. Cesno is going to adapt this system, by using special modifiers like "`unique`" to declare this variable obeys to the rule of ownership system. The challenge found here is: how to solve the conflict while the adapted basic language features are half-provided in code. A realistic problem could be "if a `unique` typed variable is assigned to a common variable, how should both variables act to keep the code understandable and reasonable". In Cesno, users have the right to write code as they want, they may not only require more code style while the CLI formatter transforms standard exchange format to what they like, they may also want some features provided by other languages. Besides the reason of language-requirement or server environment, these aspects could also make users struggling in picking languages: project targets, language features, language's strengths and weaknesses, availability of languages, whether the language is well-supported [24], [25]. Cesno does not want to be domain-specific that may cause users to learn another language when moving to a different problem, as a result, Cesno should find a proper solution of adapting those languages' features that users might want to use.

**Fifth**: To translate Cesno code into machine-based assembly language, there could be another proposed general mode of assembly language called CesAsm, which will cover the most basic instructions. CesAsm could be seen as a middle layer between highly abstracted high level languages and low level languages. Some machines might implement effective instructions like MADD in the ARM instruction set. CesAsm uses these instructions, and polyfills on the machine which it does not have (like a function that has different implementations on different machines). CesAsm will be helpful in analysing the convention between languages, since it will be designed to keep the balance between human-readable and easy to translate to machine. For the operation on basic types like `bool` or `int`, CesAsm will appear in the Cesno code to show how it converts most basic operations from high level to low level.

To make the code for design and logic cross-platform, Cesno is proposing abstract data type, which is the description of the action of the basic element. For example, a `Button` class will have different representation in HTML, Android Apps, or Qt Application; but the description about how the button looks like, how button response, or how this button is related to other elements, will be almost the same. By using these abstract types, programmers can focus on more about the design logic, let the implementation task move to the compiler, and the compiler can summon target code by their intrinsic features, or polyfilling the code.

JSON (JavaScript Object Notation) is known as a common way of serialising, storing and exchanging data. However, there are some drawbacks of JSON such as no comment permitted, or no data type information for stored strings [26]. If someone converts a `Date` object into JSON, they will get a string, but not an object containing timestamp and type information. To keep the type information and still make the data type diverse and recordable, Cesno is going to have a new format called CESON (Cesno Object Notation), which will combine the way Cesno represents object literal and format of JSON. Also, there would be support for comment, data encryption, or data type information, which could solve currently discovered problems of JSON. To enable storing objects in CESON, there will also be *object literal*, which works like a constructor, but defines the member directly in the instance (like JavaScript object literal). In order to avoid manually calculation to unnecessary members (e.g., some members could derive from other members) or create conflict (e.g. members are having conflicting members, which are not supposed to appear at the same time), this feature needs further design.

## 5 Conclusion

This article explores the potential of a C-style Object-Oriented Programming Language called Cesno. Although it is the initial stage of design, the main improvement offered by Cesno is the combination of convenient features from other languages, making it easier to learn and understand, while also reducing restrictions. To accomplish this goal, Cesno takes design ideas or functions from other languages, allows multiple forms of writing without introducing ambiguity, simplifies complex language features, and permits users to write more freely (e.g. if the user does not wish to specify a return type, they may simply leave it blank). These characteristics help to facilitate quick learning and transition, as well as aid in the writing of code with ease.

## 6 Acknowledgement

The authors gratefully acknowledge the financial support from Wenzhou-Kean University. We have the pleasure to work with people who shared a lot of suggestions, and discuss with me about the details of the language design. The list below is the GitHub account name for people who participated in the discussion (Exclude authors for this article).

**Table 1.** People Participated

| | | | |
|---|---|---|---|
| suzakuwcx | RDCarrot | langua | Cattttttttt |
| paizi | TokamakYuri | xht308 | 71e6fd52 |
| yinzhekong | LaBottle | | |

## 7 Funding

This work was supported by Student partnering with Faculty, Wenzhou-Kean University (WKUSPF2023032).

(accessed Jan. 14, 2023).